\documentstyle[12pt]{article}
\begin {document}
\title {Ghost stars in general relativity}

\author{Luis Herrera \\Instituto Universitario de F\'isica
Fundamental y Matem\'aticas, \\Universidad de Salamanca, Salamanca 37007, \\Spain \thanks{ E-mail address: lherrera@usal.es}\\Alicia Di Prisco \\Escuela de F\'{\i}sica. Facultad de Ciencias.\\ Universidad Central de Venezuela. Caracas, Venezuela. \\Justo Ospino \\Departamento de Matem\'atica Aplicada\\ Instituto Universitario de F\'isica
Fundamental y Matem\'aticas, \\Universidad de Salamanca, \\Salamanca 37007, Spain}

\date{}
\maketitle

\begin{abstract}
We explore an idea put forward  many years ago by Zeldovich and Novikov concerning the existence of compact objects endowed with arbitrarily small mass. The energy-density of such objects, which we  call ``Ghost stars'', is negative in some regions of the fluid distribution, producing a vanishing total mass. Thus, the interior   is matched on the boundary surface to Minkowski space-time. Some exact analytical solutions are exhibited and their properties are analyzed. Observational data that could confirm or dismiss the existence of this kind of stellar object is commented.
\end{abstract}

\newpage

\section{Introduction}

In their book on relativistic astrophysics, Zeldovich and Novikov (ZN) \cite{zn}  (see also \cite{z}), raise the question about the possibility of packaging the constituents of a self--gravitating fluid distribution in such a way that the total mass of the resulting compact object be arbitrarily small.

Specifically, they consider static spherically symmetric  fluid distributions, for which the line element may be written as
\begin{equation}
ds^2=e^{\nu} c^2dt^2 - e^{\lambda} dr^2 -
r^2 \left( d\theta^2 + \sin^2\theta d\phi^2 \right),
\label{metrica}
\end{equation}
where $\nu(r)$ and $\lambda(r)$ are  functions of  $r$, and $c$ is the light velocity. In this section we shall follow the notation of \cite{zn} however in  the rest of the manuscript we shall use relativistic units, in which case we put $c=G=1$.

The fluid distribution is  bounded from the exterior by a
surface $\Sigma$ whose equation is $r=r_{\Sigma}=\rm constant$.

From (\ref{metrica}) and the Einstein equations we may write
\begin{equation}
e^{-\lambda}=1-\frac{8\pi G}{r c^2}\int^{r}_ 0{ \mu r^2dr},
\label{zn1}
\end{equation}
and for the three dimensional volume element we have

\begin{equation}
dV=4\pi e^{\lambda/2}r^2 dr,
\label{zn2}
\end{equation}
where  $\mu$ denotes the energy-density of the fluid.

Then we have for the total mass(energy) the well known expression

\begin{equation}
 E=Mc^2=4\pi c^2 \int^{r_{\Sigma}}_ 0{ \mu r^2dr}.
\label{zn3}
\end{equation}

ZN also introduce the rest energy of the constituent particles $E_0$ given by

\begin{equation}
 E_0=M_0c^2= N m_0c^2,
\label{zn4}
\end{equation}
where $m_0$  is the particle mass and $N$ denotes the total number of particles which may be expressed through the particle density $n$ as 
\begin{equation}
N=  \int_ V{ n dV}.
\label{zn4b}
\end{equation}

Also,  denoting by  $E_1$ the rest energy $E_0$ plus the  kinematic energy and the  interaction energy of the constituents (excluding the gravitational interaction), we may write
\begin{equation}
 E_1=M_1c^2= c^2 \int_ V{\mu dV}=4\pi c^2 \int^{r_{\Sigma}}_ 0{ e^{\lambda/2}\mu r^2dr}.
\label{zn4}
\end{equation}

Since $e^{\lambda/2}\geq1$ then the mass defect $\Delta M=M_1-M$ should be positive.

Thus the original question posed by ZN may be rephrased as: can the constituents of a star be packaged, in such a way that the mass defect equals $M_1$?

They answer affirmatively  to the above question, and illustrate their point by analyzing the case of an ideal Fermi gas.  Although their analysis is flawed, as we shall see below, the case for the existence of stars with arbitrarily small total mass, should not be dismissed. 

Let us first reproduce the analysis of  ZN, following strictly their line of arguments (with only slight changes in notation).

Thus, let us consider an ultra-relativistic Fermi gas, characterized by an equation of state given by 
\begin{equation}
\mu=\beta n^{4/3},\qquad \beta\equiv \frac{3}{8}\hbar (3\pi^2)^{1/3}, 
\label{zn5}
\end{equation}
where $\hbar$ is the Planck constant over $2\pi$.

Next, ZN assume for the distribution of energy-density the form
\begin{equation}
\mu=\frac{a}{r^2}, \quad a=constant.
\label{zn6}
\end{equation}
It is worth emphasizing  that the above choice is justified by the fact that it coincides with the well known    Tolman $VI$ solution \cite{Tolm}, whose equation of state for large values of $\mu$ approaches that for a highly  compressed Fermi gas. 

Then using (\ref{zn6})  in (\ref{zn3}), it follows at once
\begin{equation}
M=4\pi a  r_\Sigma.
\label{zn7}
\end{equation}

On the other hand, using (\ref{zn1}),  (\ref{zn2}),  (\ref{zn4b}),  (\ref{zn5}),  (\ref{zn6}) we obtain for $N$

\begin{equation}
N=\frac{\alpha  r_\Sigma^{3/2}}{\sqrt{1-\frac{8\pi G a}{c^2}}}, \quad  \alpha\equiv \frac{8\pi}{3} \left(\frac{a}{\beta}\right)^{3/4},
\label{zn8}
\end{equation}
implying 
\begin{equation}
r_\Sigma=\alpha^{-2/3}N^{2/3}\left(1-\frac{8\pi G a}{c^2}\right)^{1/3}.
\label{zn9}
\end{equation}
Feeding back (\ref{zn9}) into (\ref{zn7}) produces

\begin{equation}
M\sim N^{2/3}\left(1-\frac{8\pi G a}{c^2}\right)^{1/3}.
\label{zn10}
\end{equation}

From (\ref{zn10}), ZN conclude that in the limit when $a\rightarrow \frac{c^2}{8\pi G}$, the total  mass $M$ tends to zero. 

Such  a conclusion is incorrect, as it is clear from (\ref{zn8}) and (\ref{zn9}) which imply that in the limit $a\rightarrow \frac{c^2}{8\pi G}$, $N$ diverges as $N^{2/3}\sim \frac{1}{(1-\frac{8\pi G a}{c^2})^{1/3}}$, thereby canceling the term $(1-\frac{8\pi G a}{c^2})^{1/3}$ in (\ref{zn10}). This is also evident from (\ref{zn7}) which  shows that $M$  does not tend to zero for any value of $a$ (different from zero).

In general, it should  be clear from its very definition (\ref{zn3}), that $M$ cannot be zero for any positive defined energy-density function $\mu$. Thus,  vanishing total mass is only possible if we accept  the existence  of fluid distributions allowing negative-energy density, or in the trivial case $\mu=0$.

The appearance of negative energy-density (mass)  in general relativity has been considered in the past by several researchers, starting with a paper by Bondi \cite{bon}. This issue also  appears in relation with the Reissner-Nordstrom solution and classical electron models  (see \cite{co1,bo1,pap,neg} and references therein). More recently, negative masses have been invoked in the construction of some cosmological models (see \cite{nm,nm1} and references  therein). Also, it is worth mentioning that negative energy-density appears in hyperbolically symmetric fluids (see \cite{negh,hh} and references therein).  In all the cases above,  quantum effects were not taken into account. However, in spite of these examples, we believe that it is fair to say that the assumption of positive energy-density  is well justified,  at the classic level, for any realistic fluid. 

Notwithstanding,   the situation is quite different in the quantum regime. Indeed, as it has been argued in recent past  (see \cite{we1,we2,we3,cqg,pav} and references therein), the appearance of negative energy-density  is possible, whenever quantum effects are expected to be relevant.

Thus, the  idea of compact objects with arbitrarily small total mass is still  feasible, if we accept the possibility of  negative energy-density. We call such objects ``Ghost stars'' , in analogy with a somehow similar situation observed in some Einstein--Dirac  neutrinos (named  ghost neutrinos)  which do not produce gravitational field but  still are characterized by  non vanishing current density \cite{gn,gn1,gn2}.

 In this work we shall explore such a possibility, by presenting explicit analytical models of  ghost stars.
\section{The Einstein equations for static locally anisotropic fluids}
In what follows we shall briefly summarize the definitions and main equations required for describing spherically symmetric static anisotropic fluids. We shall heavily rely on \cite{146}, therefore we shall omit many steps in the calculations, details of which the reader may  find in \cite{146}.

We consider a spherically symmetric distribution  of static 
fluid,   bounded by a spherical surface $\Sigma$. The fluid is
assumed to be locally anisotropic (principal stresses unequal).

The justification to consider anisotropic fluids, instead of isotropic ones,  is provided by the fact that pressure anisotropy is produced by many different physical phenomena of the kind expected in  gravitational collapse scenario (see \cite{report} and references therein). In particular  we expect that the final stages of stellar evolution should be accompanied by intense dissipative processes, which, as shown in \cite{ps}, should produce pressure anisotropy.

In curvature coordinates (using relativistic units) the line element reads (please  notice that we are using  signature $-2$, instead $+2$ as  in \cite{146})
\begin{equation}
ds^2=e^{\nu (r)} dt^2-e^{\lambda (r)}
dr^2-r^2(d\theta^2+r^2sin^2\theta d\phi^2),  \label{metric}
\end{equation}

 \noindent which has to satisfy the Einstein equations. For  a locally anisotropic fluid they are
\begin{equation}
8\pi \mu=\frac{1}{r^2}-e^{-\lambda}
\left(\frac{1}{r^2}-\frac{\lambda'}{r} \right), \label{fieq00}
\end{equation}

\begin{equation}
8\pi  P_r =-\frac{1}{r^2} +e^{-\lambda}
\left(\frac{1}{r^2}+\frac{\nu'}{r}\right), \label{fieq11}
\end{equation}

\begin{eqnarray}
8\pi P_\bot = \frac{e^{-\lambda}}{4} \left(2\nu''+\nu'^2 -
\lambda'\nu' + 2\frac{\nu' - \lambda'}{r}\right), \label{fieq2233}
\end{eqnarray}
where primes denote derivative with respect to $r$, and $\mu, P_r$ and $P_\bot$ are  proper  energy-density,  radial pressure and  tangential pressure respectively.

The above is  a system of three ordinary differential equations for the five  unknown functions $\nu, \lambda, \mu, P_r, P_\bot$, accordingly their  solutions would depend on two arbitrary functions.

From the above field equations, the Tolman--Oppenheimer--Volkof  equation follows
\begin{equation}
P'_r=-\frac{(m + 4 \pi P_r r^3)}{r \left(r - 2m\right)}\left(\mu+P_r\right)+\frac{2\left(P_\bot-P_r\right)}{r},\label{ntov}
\end{equation}
where we have introduced the mass function  $m$ \cite{Misner} defined by 
\begin{equation}
e^{-\lambda}=1-\frac{2m(r)}{r}.
\label{m}
\end{equation}

In \cite{146} a general algorithm to  express any solution for anisotropic fluids in terms of two generating functions was proposed (see also \cite{Lake2}). It generalizes a previous work by Lake for isotropic fluids \cite{KLake}.

Specifically it was shown that the general  line element corresponding to any solution to the system (\ref{fieq00})--(\ref{fieq2233}) may be written as

\begin{eqnarray}
ds^2&=&e^{\int (2z(r)-2/r)dr}dt^2\nonumber \\&-&\frac{z^2(r) e^{\int(\frac{4}{r^2
z(r)}+2z(r))dr}} {r^6(-2\int\frac{z(r)(1+\Pi (r)r^2)
e^{\int(\frac{4}{r^2
z(r)}+2z(r))dr}}{r^8}dr+C)}dr^2\nonumber \\&-&r^2d\theta^2-r^2sin^2\theta
d\phi^2. \label{metric2b}
\end{eqnarray}

\noindent with $\Pi(r)=8\pi (P_r-P_\bot),$ and 
\begin{equation}
e^{\nu (r)}=e^{\int (2z(r)-2/r)dr} 
\label{v1b}
\end{equation}
where  $C$ is a constant of integration.

The physical variables may be written as 
\begin{equation}
4\pi P_r=\frac{z(r-2m)+m/r-1}{r^2}\label{Pr},
\end{equation}
\noindent 

 \begin{equation}
4\pi \mu =\frac{m^{\prime}}{r^2}\label{rho},
\end{equation}
and 
\begin{equation}
4\pi P_\bot=\left(1-\frac{2m}{r}\right)\left(z^{\prime}+z^2-\frac{z}{r}+\frac{1}{r^2}\right)+z\left(\frac{m}{r^2}-\frac{m^{\prime}}{r}\right).
\label{Pbot}
\end{equation}

In order to match smoothly the  metric  (\ref{metric}) with the Schwarzschild metric on the boundary surface
$r=r_\Sigma=constant$, we  require the continuity of the first and the second fundamental
forms across that surface, producing 
\begin{equation}
e^{\nu_\Sigma}=1-\frac{2M}{r_\Sigma},
\label{enusigma}
\end{equation}
\begin{equation}
e^{-\lambda_\Sigma}=1-\frac{2M}{r_\Sigma},
\label{elambdasigma}
\end{equation}
\begin{equation}
\left[P_r\right]_\Sigma=0,
\label{PQ}
\end{equation}
where  subscript $\Sigma$ indicates that the quantity is
evaluated on the boundary surface $\Sigma$.

The above conditions  hold for any value of $M$,  including $M=0$. 

For configurations with $M=0$ we obtain from (\ref{Pr}) and (\ref{PQ})
\begin{equation}
z_\Sigma=\frac{1}{r_\Sigma}.
\label{zb}
\end{equation}

We shall next present  solutions describing fluid spheres with vanishing total mass. For doing that we shall resort to a variety of assumptions, some of which are usually invoked in the modeling of relativistic stars. 

\section{Conformally flat  ghost stars}

If we assume the space-time within the fluid distribution to be conformally flat, then the two generating functions  read \cite{j1}
\begin{equation}
z=\frac{2}{r}\pm \frac{e^{\frac{\lambda}{2}}}{r}\, tanh\left(\int
\frac{e^{\frac{\lambda}{2}}}{r}dr\right).\label{zlambda}
\end{equation}
\noindent and
\begin{equation}
\Pi=r\left(\frac{1-e^{-\lambda}}{r^2}\right)^\prime.
\label{Pilambda}
\end{equation}
In (\ref{zlambda}) we shall choose the minus sign, since the plus sign leads (in this case) to a model not satisfying the boundary condition (\ref{zb}).

We shall present two conformally flat models of ghost star. For  that purpose we shall complement the conformal flatness condition with some additional restrictions.

\subsection{Ghost star with a given density profile}

Let  us assume a density profile of the form 
\begin{equation}
4\pi\mu=\sum^n_{i=0}{a_ i r^{i-2}},
\label{p1}
\end{equation}
which using (\ref{rho}) produces
\begin{equation}
m=\sum^n_{i=0}{\frac{a_ i }{i+1}r^{i+1}}.
\label{p2}
\end{equation}

Since the total mass is assumed to vanish then the following condition has to be satisfied

\begin{equation}
\sum^n_{i=0}{\frac{\bar a_ i }{i+1}}=0, 
\label{p3}
\end{equation}
with $\bar a_i=a_i r^i_\Sigma$.

In order to describe a specific model, let us restrict the expression (\ref{p1})  to $n=2$.

Thus we obtain for the energy-density and the mass function
\begin{equation}
4\pi \mu=-\frac{3}{2r^2}+\frac{a_1}{r}+a_2,
\label{p4}
\end{equation}
and
\begin{equation}
m=-\frac{3}{2}r+\frac{a_1}{2}r^2+\frac{a_2}{3}r^3,
\label{p5}
\end{equation}
where  we have chosen $a_0=-\frac{3}{2}$ to simplify the calculation of the second term on the right of (\ref{zlambda}).

Then the condition (\ref{p3}) reads
\begin{equation}
a_2=\frac{9}{2r^2_\Sigma}-\frac{3 a_1}{2 r_\Sigma},
\label{p6}
\end{equation}

and using (\ref{p5}), (\ref{p6}) in (\ref{m}) we obtain

\begin{equation}
e^{-\lambda}=4-\frac{3r^2}{r^2_\Sigma}- a_1 r\left(1-\frac{r}{ r_\Sigma}\right).
\label{p7}
\end{equation}

\noindent With the expression for  $\lambda$ given by   (\ref{p7}),   the two generating functions for this case become
\begin{equation}
z=\frac{5}{2r}-\sqrt{\frac{a_2}{-24+6 a_1r+4a_2r^2}},
\label{pz}
\end{equation}
and
\begin{equation}
\Pi=\frac{6}{r^2}-\frac{a_1}{r}.
\label{Pz}
\end{equation}

The constant $a_1$ may be easily obtained from (\ref{pz}), using condition (\ref{zb}), it reads
\begin{equation}
a_1=\frac{12}{r_\Sigma},
\label{a1}
\end{equation}
which combined with (\ref{p6}) produces
\begin{equation}
a_2=-\frac{27}{2r^2_\Sigma}.
\label{a1}
\end{equation}
With the two expressions above we finally obtain for $z$ and $m$
\begin{equation}
z=\frac{6r -5 r_\Sigma}{r(3r-2r_\Sigma)},
\label{zf}
\end{equation}
\begin{equation}
m=-\frac{3}{2}r+\frac{6 r^2}{r_\Sigma}-\frac{9 r^3}{2r^2_\Sigma},
\label{p5b}
\end{equation}

and using using (\ref{Pr}), (\ref{p4}) and (\ref{Pz}) we obtain for the energy-density, the radial pressure, and $\Pi$
\begin{equation}
4\pi \mu=-\frac{3}{2r^2}+\frac{12}{r_\Sigma r}-\frac{27}{2r^2_\Sigma},
\label{p4n}
\end{equation}

\begin{eqnarray}
4\pi P_r=\frac{27}{2r^2_\Sigma}-\frac{21}{r r_\Sigma}+\frac{15}{2r^2},
\label{prcf}
\end{eqnarray}

\begin{equation}
\Pi=\frac{6}{r^2}-\frac{12}{r r_\Sigma}.
\label{Pzn}
\end{equation}

Using (\ref{p4n} ) and (\ref{Pzn}), the reader can easily check that the condition of conformal flatness (see Eq.(29) in \cite{c1})
\begin{equation}
P_r-P_\bot=\frac{1}{r^3}\int^r_0{r^3 \mu^\prime dr},
\label{ccf}
\end{equation}
is satisfied.

From (\ref{p4n}) we see that $\mu$ is negative in the intervals $0< r <0.15 r_\Sigma$ and $r> 0.73r_\Sigma$. As it is apparent from the expressions of the physical variables, the fluid distribution has a singularity at the origin ($r=0$) and therefore the center should be excluded from the discussion. The best  way to handle this drawback consists in assuming that a vacuum cavity surrounds the center. Denoting the equation of the boundary of the cavity by $r=r_i=constant$, we obtain from (\ref{p5b}) $r_i=\frac{r_\Sigma}{3}$, which ensures the continuity of the mass function on that surface. However the radial pressure is discontinuous on that surface, and therefore it is a thin shell.

\subsection{Ghost star with the Gokhroo and Mehra ansatz}
We shall now complement the conformal flatness condition with an ansatz  proposed by  Gokhroo and Mehra \cite{22}, which leads to physically satisfactory models for compact objects.

Thus we shall assume for  $\lambda$ the condition

\begin{equation}
e^{-\lambda}=1-\alpha r^2+\frac{3K\alpha r^4}{5r_\Sigma^2},
\label{lgm}
\end{equation}
producing, because of (\ref{fieq00}) and (\ref{m})
\begin{equation}
\mu=\mu_0\left(1-\frac{Kr^2}{r_\Sigma^2}\right),
\label{mugm}
\end{equation}
and
\begin{equation}
m(r)=\frac{4\pi\mu_0r^3}{3}\left(1-\frac{3Kr^2}{5r_\Sigma^2}\right),
\label{mgm}
\end{equation}

where $K$ is a constant, $\mu_0$ is the central density and 

\begin{equation}
\alpha\equiv \frac{8 \pi \mu_0}{3}.
\label{gmn1}
\end{equation}

Since we must impose $m(r_\Sigma)=0$ then $K=\frac{5}{3}$.

\noindent Feeding back this value of $K$ into (\ref{lgm}), (\ref{mugm}), and (\ref{mgm}) we obtain 
\begin{equation}
    4\pi \mu=\frac{6}{r^2_\Sigma}\left ( 1-\frac{5r^2}{3r^2_\Sigma}  \right ),\label{mucfG}
\end{equation}

\begin{equation}
    m=\frac{2r^3}{r^2_\Sigma}\left ( 1-\frac{r^2}{r^2_\Sigma}  \right ),\label{mcfG}
\end{equation}
and 

\begin{equation}
e^{-\lambda}=1-\frac{4 r^2}{r^2_\Sigma}+\frac{4r^4}{r^4_\Sigma},
\label{mmcfG}
\end{equation}
where we have chosen $\alpha=\frac{4}{r^2_\Sigma}$, in order to facilitate the calculation of the second term on the right of (\ref{zlambda}).
Thus, we obtain for $z$
\begin{equation}
    z=\frac{3}{r}-\frac{2r}{2r^2-r^2_\Sigma},\label{zcfG}
\end{equation}
whereas for $\Pi$ we obtain from 
(\ref{Pilambda})
\begin{equation}
     \Pi=-\frac{8r^2}{r^4_\Sigma},
\end{equation}
and from  (\ref{Pr}) we obtain the expression for $P_r$
 \begin{equation}
    8\pi P_r=\frac{4}{r^2}\left (1-\frac{4r^2}{r^2_\Sigma}+\frac{3r^4}{r^4_\Sigma}\right).
 \end{equation}

As it follows from (\ref{mucfG}) the energy-density becomes negative for  $r> 0.77r_\Sigma$.

As in the precedent model, there appears a singularity at the center which could be embedded in a vacuum cavity bounded by a thin shell.

\section{Ghost stars with vanishing complexity factor}
The complexity factor is a scalar function intended to measure the degree of complexity of a given fluid distribution, it was introduced in \cite{c1} for static spherically symmetric configurations.

The vanishing complexity factor condition reads (see \cite{c1} for details), 

\begin{equation} 
 \Pi=\frac{4\pi}{r^3} \int^r_0{\tilde r^3 \mu' d\tilde r},
\label{vcfc}
\end{equation}
please notice that the symbol $\Pi$ here differs from the one in \cite{c1} by a factor $8 \pi$.
 
 \noindent Using  (\ref{Pr})-(\ref{Pbot}) and  (\ref{vcfc}) we are led to the following differential equation for $z$
 \begin{eqnarray}
     &&2\left(1-\frac{2m}{r}\right)\left(z^\prime+z^2\right)-\left(\frac{2}{r}-\frac{5m}{r^2}+\frac{m^\prime}{r}\right)\left(2z-\frac{1}{r}\right)\nonumber \\&+&\frac{2}{r}-\frac{4m}{r^3}=0.
     \label{ecz}
 \end{eqnarray}

 The first integral of the above equation (for $m$) reads
\begin{equation}
    1-\frac{2m}{r}=e^{\int _r ^ {r_\Sigma} \frac{4(r^2 z^\prime+r^2z^2-2r z+2)}{2r^2z-r}dr },\label{Inz}
\end{equation}
 \noindent from which we see that for any $z$ satisfying (\ref{zb}) we have a model with  vanishing complexity factor.  However, we shall follow here a different strategy, and we  shall  present two models of ghost star satisfying the vanishing complexity factor condition, by imposing two different additional restrictions.

\subsection{A model with a given energy-density profile}
 In order to specify this model we shall propose the following energy--density  profile, 
 \begin{equation}
   8\pi \mu=\frac{1-9(\frac{r}{r_\Sigma})^8}{r^2}, \label{mufc} 
 \end{equation}
 \noindent  producing for $m$
 \begin{equation}
    m=\frac{r}{2}\left[1-\left(\frac{r}{r_\Sigma}\right)^8\right],\label{mfc}
 \end{equation}
the reason behind this choice is simply that it allows the integration of (\ref{ecz}).

 \noindent Indeed, feeding back  (\ref{mfc}) into  (\ref{ecz}), we may easily integrate this equation for  $z$, obtaining
\begin{equation}
    z=\frac{1}{c_1 r^2-r},\label{solz}
\end{equation}
\noindent where $c_1$ is a constant of integration, which according to (\ref{zb}) reads

\begin{equation}
c_1=\frac{2}{r_\Sigma}.
\label{zb1}
\end{equation}

Having obtained the two ¡generators of the solution we may write for $P_r$ and $P_\bot$
 \begin{eqnarray}
     8\pi P_r&=&-\frac{1}{r^2}+\frac{r^6}{r^8_\Sigma}\left ( \frac{3-\frac{2r}{r_\Sigma}}{\frac{ 2r}{r_\Sigma} -1} \right ),\\
     8\pi P_\bot &=&\frac{4r^7}{r^9_\Sigma \left(\frac{ 2r}{r_\Sigma} -1\right)}.
 \end{eqnarray}

 In this model the energy density becomes negative for values of $r$ in the interval $[.76 r_\Sigma<r,r=r_\Sigma]$. As in the previous model, the fluid presents a singularity at the origin, which could be surrounded by a cavity bounded by a thin shell.

 \subsection{Ghost star with vanishing active gravitational mass}

For this model we shall additionally assume that the active gravitational (Tolman) mass \cite{agm}  vanishes.

This last condition implies (see eq.(7.30) in \cite{report})
\begin{equation}
m+4\pi P_r r^3=0.
\label{am1}
\end{equation}
Feeding the above condition into (\ref{ntov}) and using (\ref{vcfc}), we obtain 
\begin{equation}
P^\prime_r+\frac{\Pi}{4\pi r}=0,
\end{equation}
which can be easily transformed  into
\begin{equation}
P^{\prime \prime}_r+\frac{4 P^\prime_r}{r}+\frac{\mu^\prime}{r}=0.
\label{am2}
\end{equation}

In order to find a solution to the above equation, we shall split it in two equations, as
\begin{equation}
P^{\prime \prime}_r+\frac{3 P^\prime_r}{r}=0,
\label{am2s}
\end{equation}
and 
\begin{equation}
\frac{P^\prime_r}{r}+\frac{\mu^\prime}{r}=0,
\label{am2s1}
\end{equation}
whose solutions reads
\begin{equation}
P_r=b\left(\frac{1}{r^2}-\frac{1}{r^2_\Sigma} \right),
\label{am3}
\end{equation}
and
\begin{equation}
\mu=b\left(\frac{3}{r_\Sigma^2} -\frac{1}{r^2}\right),
\label{am7}
\end{equation}
where boundary conditions have been used, and  $b$ is a  constant of integration.

Using (\ref{am3})  in (\ref{am1}), we obtain for the mass
\begin{equation}
m=4\pi r^3 b\left(\frac{1}{r_\Sigma^2} - \frac{1}{r^2}\right),
\label{am6}
\end{equation}
while using (\ref{am7}) in (\ref{vcfc}) we obtain for $\Pi$
\begin{equation}
\Pi=\frac{8\pi b}{r^2}.
\label{am8}
\end{equation}

In this model the energy density becomes negative in  the interval   $0<r< 0.58 r_\Sigma$ (if we assume $b>0$). 
As in the previous  models the physical variables exhibit a singular behavior at the center, and any surface delimiting a vacuum cavity surrounding the center would be a thin shell.

\section{Discussion}

Exploring the possibility of the  existence of compact objects   endowed with vanishing total mass(energy), we have  presented four exact solutions to Einstein equations for static spherical distribution of anisotropic fluids, sharing this property.  Such solutions must, within some regions of the distribution, be endowed with  negative energy-density. Negative energy-density values appear indistinctly in  outer or in  inner regions, depending on the model,    not a  universal pattern of distribution having been  detected.

Although some of the assumptions adopted to obtain the presented solutions (e.g. the vanishing complexity factor or the conformal flatness) are physically meaningful, the obtained solutions are intended only to  illustrate the above mentioned possibility but not  to describe any specific astrophysical scenario. A pending problem regarding this issue consists in finding exact solutions for ghost stars,  directly related to relevant astrophysical data.

In the same order of ideas, an important open question concerning ghost stars is related with  possible astrophysical observations that  could confirm (or dismiss) the existence of this kind of object. We have in mind  for example a  new trend of investigations   based on  the recent observations of shadow images of the gravitationally collapsed objects at the center of the elliptical galaxy $M87$  and at the center of the Milky Way galaxy by the Event Horizon Telescope (EHT) Collaboration (see \cite{et1, et2,et3,et4} and references therein). More specifically, we wonder if it could be possible to establish the existence of a ghost star by its shadow.

The solutions we have presented should be considered as the final state of collapsing stars, where quantum effects become relevant during the evolution process. Accordingly it is of utmost interest to describe the process leading to final stage with vanishing total mass. In other words we should be able to provide exact solutions describing such process. This problem is out of the scope of this manuscript, but remains one of the most relevant questions to solve, concerning the physical viability of ghost stars.

Regarding the formation of a ghost star,  it should be clear  from elementary physical considerations that, as a final  product of gravitational collapse, the  formation  of such configurations must be preceded by an intense radiative process. The problem about the efficiency   of energy release in gravitational collapse has been discussed by several authors (see \cite{dys,is,mi} and references therein).
Some of these authors conclude that a $100\%$ efficiency (all the mass is radiated away)  is possible under rather mild restrictions \cite{dys,mi}, while others \cite{is} claim that  $100\%$ efficiency is forbidden under physically meaningful  conditions, among which positive energy-density plays a relevant role.  Thus, the violation of  such a condition, as it happens in our models,  is a strong argument to believe that $100\%$ efficiency could   be a likely possibility. In such a case, the detection  of a strong  emission of radiation, might  indicate the location of a ghost star.

We would like to conclude with five  remarks oriented to encourage future research on this issue
\begin{itemize}
\item We have explored the possibility of ghost stars within the context of general relativity. It would be interesting to explore such a possibility under some of the extended theories  of gravity \cite{ext}.
\item For reasons exposed before, we have considered anisotropic fluids. However it seems clear that ghost stars models described  by isotropic fluids should also  exist. It could be interesting to find some  models of this kind.
\item All the models here presented exhibit  a singularity at the origin. In order to exclude such region we have proposed to surround the center by a vacuum cavity. However, in all examples analyzed  the boundary surface of such cavity appears to be a thin shell. It would be interesting to find singularity-free solutions, or  singular solutions whose center could be embedded in a vacuum cavity delimited  by a regular boundary.
\item We would like to insist on the importance to find exact (analytical or numerical) solutions describing the evolution leading to a ghost star. 
\item Alternatively, it could be also of interest to find solutions describing the evolution of an initial ghost star leading  to a $M>0$ object, by absorbing radiation. As strange as this scenario might look like (compact object absorbing radiation), it is worth noticing that it  has been invoked in the past to explain the origin of gas in quasars \cite{mat}.  A semi-numerical example for such a model is described in \cite{eh}.
\end{itemize}

\end{document}